\documentclass [a4paper,fleqn]{article}

\usepackage {amsmath}
\usepackage{amssymb}
\usepackage{cite}
\numberwithin{equation}{section}
\numberwithin{table}{section}
\begin{document}

\title{Exact solutions for one of the extensive chaos model}
\author{Nikolai A. Kudryashov}
\date{Department of Applied Mathematics\\
Moscow  Engineering and Physics Institute\\
(State university)\\
31 Kashirskoe Shosse,  115409\\
Moscow, Russian Federation} \maketitle

\begin{abstract}
Sampling equation method is presented to look for exact solutions of nonlinear
differential equations. Application of this approach to one of the extensive chaos model
is considered. Exact solutions of this model in travelling wave are given. Nonlinear
evolution equation for the considered extensive chaos model is shown to have solitary and
periodical waves.
\end{abstract}



\section{Introduction}
In recent years one can observe a systematical study of a novel type of chaos that is
called by "soft-mode turbulence" \cite{1,2,3}. These chaos types are characterized by a
smooth interplay of different spatial scales. Properties of these types are qualitatively
different from the well known models that are described by the complex Ginzburg--Landau
and the Kuramoto--Sivashinsky equations.

The simplest model exhibiting the soft-mode turbulence can be described by the higher
order nonlinear evolution equation with the simplest nonlinearity.

This equation was introduced by N.A. Kudryashov \cite{4} and V.N. Nikolaevskiy \cite{5}
to describe longitudinal seismic waves in viscoelastic media. The simplest case of this
equation takes the form

\begin{equation}
\label{1.1}u_t + uu_x +\beta u_{xx} +\delta u_{xxxx} +\varepsilon u_{xxxxxx}=0
\end{equation}

It is known that the Ginzburg--Landau and the Kuramoto--Sivashinskiy equations are not
integrable equations because these ones do not pass the Painleve test \cite{6,7}. However
these equations have some list of special solutions \cite{6,7,8,9,10,11,12}.

Eq. \eqref{1.1} can be normalized. Assuming $\varepsilon \neq 0, \delta \neq 0$ and
setting

\begin{equation}
\label{1.2}u=\frac{\delta^2}\varepsilon \left(\frac\delta\varepsilon\right)^{\frac12}
u',\qquad x=\left(\frac\varepsilon\delta\right)^{\frac12} x',\qquad
t=\frac{\varepsilon^2}{\delta^3} t',\qquad \sigma=\frac{\beta\varepsilon}{\delta^2}
\end{equation}

Then Eq.\eqref{1.1} takes the form

\begin{equation}
\label{1.3}u_t +uu_x +\sigma u_{xx} + u_{xxxx} +u_{xxxxxx}=0
\end{equation}
(the primes of the variables are omitted).

Equation \eqref{1.3} is invariant under transformations

\begin{equation}
\label{1.4}u\rightarrow -u, \qquad x\rightarrow -x
\end{equation}
which allows us to study this equation for $x\geq 0$.

Eq.\eqref{1.1} does not pass the Painlev\'e test and this is not integrable equation but
one can expect that Eq.\eqref{1.1} has some special solutions.

The aim of this letter is to present some exact solutions of Eq.\eqref{1.1}. The outline
of this letter is as follows. The sampling equation method to look for exact solutions of
nonlinear differential equations is discussed in Section 2. Application of this approach
to search exact solitary solutions of Eq.\eqref{1.1}  is considered in Section 3. Exact
periodic solutions  are presented in Section 4.

\section{Sampling equation method}
It is well known that all nonlinear differential equations can be connectionally divided
into three types: exactly solvable, partially solvable and those that have no exact
solution.
 At the present we have a lot of different approaches to look for exact
solutions of nonlinear differential equations (see, for a example, refs.
\cite{6,7,8,9,10,11,12,13,14,15,16,17,18,19,20}). Usually investigators use some sampling
functions that are hyperbolic and elliptic functions. However one can note that as a rule
partially solvable nonlinear equations have exact solutions that are general solutions of
solvable equation of lesser order.
 In this connection we apply later the sampling
equation method to look for exact solutions of Eq.\eqref{1.1}. Our approach takes into
consideration the following simple idea.

Let us assume we have two differential equations

\begin{equation}
\label{2.1}E[y]=0
\end{equation}
and
\begin{equation}
\label{2.2}D[u]=0
\end{equation}
and let us also assume that Eq.\eqref{2.1} is not integrable equation but Eq. \eqref{2.2}
is solvable equation of lesser order then Eq.\eqref{2.1}. If we find the transformation
for solution $y$ of Eq.\eqref{2.1} that allows us to connect $y$ with the general
solution of Eq.\eqref{2.2} we have the following relation between Eq.\eqref{2.1} and
\eqref{2.2}

\begin{equation}
\label{2.3}E[y] =\Hat{A}D[u]
\end{equation}
where $\hat{A}$ is a operator and $y$ is a transformation that is determined by the
formula

\begin{equation}
\label{2.4}y=F(u)
\end{equation}

This raises the question as to whether finding transformation \eqref{2.4} and exactly
solvable equation \eqref{2.2} as the sampling equation.

One of the impressive method to look for the transformation like \eqref{2.4} is the
singular manifold method by J.Weiss, M.Tabor and G.Carnevalle \cite{21} that is used to
study both integrable and nonintegrable differential equations. Success of this approach
for nonintegrable differential equations is explained by so-called truncated expansions
that are transformations similar to formula \eqref{2.4}. In this case for the polynomial
class of nonintegrable equations \eqref{2.1} one can suggest corresponding exactly
solvable equation \eqref{2.2} as the Riccati equation, the elliptic equation or other
solvable ordinary differential equation \cite{6,7,10,22}.

 As a example let us consider the
ordinary differential equation in the form

\begin{equation}
\label{2.5}E[y] = y_{xxxx} + yy_{xxx} - 6 yy_{xx} -6y^2_x - 6y^2y_x -\beta y=0
\end{equation}

This equation is not integrable equation but this one has some exact solutions.

Taking into consideration leading members of Eq.\eqref{2.5} one can find that solution of
Eq.\eqref{2.5} have the second degree singularity. In this connection we can find
solution of Eq.\eqref{2.5} using the truncated expansion

\begin{equation}
\label{2.6}y(z) =A_0 + A_1 Y + A_2 Y^2
\end{equation}
where $A_0,A_1$ and $A_2$ are unknown parameters and $Y(z)$ satisfies to the Riccati
equation

\begin{equation}
\label{2.7}D[Y]= Y_z +Y^2 -\alpha=0
\end{equation}

Here $\alpha$ is a parameter that will be found too. Substituting transformation
\eqref{2.6} info Eq.\eqref{2.5} and taking into account Eq.\eqref{2.7} and its
consequences

\begin{equation}
\label{2.8}Y_{zz} =2Y^3 -2\alpha Y
\end{equation}

\begin{equation}
\label{2.9}Y_{zzz} =-6Y^4 + 8\alpha Y^2 -2\alpha^2
\end{equation}

\begin{equation}
\label{2.10}Y_{zzzz}=24Y^5 -40 \alpha Y^3 +16\alpha^2 Y
\end{equation}
we have solution of Eq.\eqref{2.5} at $\beta=0$ that is expressed by Eq.\eqref{2.6} where
$Y(z)$ is the solution of Eq.\eqref{2.7}

One can use another transformation

\begin{equation}
\label{2.11}y(z) =B_0 + B_1 R
\end{equation}

Where $B_0$ and $B_1$ are constant that are found. As this takes place we take into
consideration that $R$ has second degree singularity and $R=R(z)$ is a solution of the
elliptic function equation

\begin{equation}
\label{2.12}R_z^2=-2R^3 + aR^2 +2 bR +d
\end{equation}
In this case we obtain the elliptic solution at $\beta=0$ again.

However if we use formula \eqref{2.11} and take  the first Painlev\'e equation except
Eq.\eqref{2.12}

\begin{equation}
\label{2.13}R_{zz} =3R^2 +\beta x
\end{equation}
we find that Eq.\eqref{2.5} has exact solution \eqref{2.11} at $B_0=0, B_1=1$ and
$\beta\neq0$ where $R(z)$ is the Painlev\'e transcendent. We can see that we have
obtained much more interesting solution of Eq.\eqref{2.5} than \eqref{2.7}and
\eqref{2.11}. This solution can not be found using sampling functions.

\section{Exact solitary solutions of Eq.(1.1).}

Let us look for exact solutions of Eq.\eqref{1.1} in the form of travelling waves using
variables

\begin{equation}
\label{3.1}u(x,t)=y(z), \qquad z=x-C_0 t
\end{equation}

Eq.\eqref{2.1} takes the form after integration over $z$

\begin{equation}
\label{3.2}C_1 -C_0 y +\frac12 y^2 +\beta y_z +\delta y_{zzz} + \varepsilon y_{zzzzz}=0
\end{equation}

Assuming

\begin{equation}
\label{3.3}y=a_0z^p
\end{equation}
and substituting into leading members of Eq.\eqref{3.2} we have $a_0=30240\varepsilon$
and $p=-5$. Solution of Eq.\eqref{3.2} has fifth degree singularity and following to the
sampling equation method  we can look for the exact solution of Eq.\eqref{3.2} in the
form

\begin{equation}
\label{3.4}y (z) = A_0+ A_1 Y  +A_2
  Y  ^{2}+ A_3  Y
 ^{3}+A_4 Y  ^ {4}+A_5  Y  ^{5}
\end{equation}

Where $Y(z)$ satisfies to the Riccati equation

\begin{equation}
\label{3.5}D[Y]=Y_z +Y^2 -\alpha =0
\end{equation}

Constants $A_0, \,A_1, \,A_2, \,A_3, \, A_4, \,A_5$ and $\alpha$ are found after
substitution of the truncated expansion \eqref{3.4} into Eq.\eqref{2.1}. We need also to
take into account the following formulas

\begin{equation}
\begin{gathered}
\label{3.6}
Y_{zz}=2Y^3 -2\alpha Y\\ \\
Y_{zzz} =-6Y^4 +8\alpha Y^2 -2\alpha^2\\ \\
Y_{zzzz}=24Y^5 -40\alpha Y^3 +16\alpha^2 Y\\ \\
Y_{zzzzz} =-120Y^6 +240\alpha Y^4 -136\alpha^2Y^2 +16\alpha^3
\end{gathered}
\end{equation}

As a result of calculations we have

\begin{equation}
\begin{gathered}
\label{3.7}A_5=30240\,\varepsilon, \quad A_4=0, \quad A_3=\frac
{2520\delta}{11}-50400\varepsilon \alpha, \quad A_2=0,\\ \\
 A_1=-{\frac
{2520}{11}}\,\delta\,\alpha+20160\,\varepsilon\,{\alpha} ^{2}+{\frac
{1260}{251}}\,\beta-{\frac {12600}{30371}}\,{\frac {{ \delta}^{2}}{\varepsilon}}, \quad
A_0=C_0
\end{gathered}
\end{equation}

Where $\beta$ is determined by the formula

\begin{equation}
\label{3.8}\beta=-{\frac {213811840\,{\varepsilon}^{3}{\alpha}^{3}
-10204656\,\delta\,{\varepsilon}^{2}{\alpha}^{2}-2045\,{\delta}^{3}-92400
\,{\delta}^{2}\varepsilon\,\alpha}{121 \varepsilon\, \left( 9240\,\varepsilon\,
\alpha+79\,\delta \right) }}
\end{equation}

Denoting

\begin{equation}
\label{3.9}\alpha=\frac{\delta w}\varepsilon
\end{equation}
we obtain for $w$ the following six values

\begin{equation}
\begin{gathered}
\label{3.10} w_1=-{\frac {1}{220}}, \quad w_2=-{\frac {5}{176}}, \quad w_3=-{\frac
{1}{440}}
\end{gathered}\end{equation}

\begin{equation}
\begin{gathered}
\label{3.11}w_4 =\frac1{52800} \left(557 - \frac{46031}{m} + m\right)
\end{gathered}\end{equation}

\begin{equation}
\begin{gathered}
\label{3.12}m=(113816753 +1260 \sqrt{8221079733})^\frac13 \approx 610,966
\end{gathered}\end{equation}

\begin{equation}
\begin{gathered}
\label{3.13}w_{5,6} =\frac1{52800} \left(\frac{46031}{2m} - \frac m2 + 557 \pm
\frac{i\sqrt{3}}2 \left( m+ \frac{46031}{m}\right) \right)
\end{gathered}
\end{equation}

Exact solutions of Eq.\eqref{1.1} can be written in the form

\begin{equation}
\begin{gathered}
\label{3.14}y \left( z \right) =30240\,\varepsilon\, Y ^{5}+ \left( {\frac
{2520}{11}}\,\delta-50400\,\varepsilon\,
\alpha \right)  Y^{3}+\\
\\+ \left( -{ \frac
{2520}{11}}\,\delta\,\alpha+20160\,\varepsilon\,{\alpha}^{2}+{ \frac
{1260}{251}}\,\beta-{\frac {12600}{30371}}\,{\frac {{\delta}^{2} }{\varepsilon}}\right)
Y +C_0
\end{gathered}\end{equation}
where $Y=Y(z)$ is a solution of Eq.\eqref{3.5}

\begin{equation}
\label{3.15}Y(z)=\sqrt{\alpha}\tanh \left(\sqrt{\alpha} z +\varphi_0\right)
\end{equation}

 Constant $C_1$ is determined by formula

\begin{equation}\begin{gathered}
\label{3.16} C_1={\frac {4112640}{11}}\,{\frac {{\delta}^{5}{w}^{4}}{{\varepsilon
}^{3}}}-9999360\,{\frac {{\delta}^{5}{w}^{5}}{{\varepsilon}^{3}}}-{\frac
{5080320}{251}}\,{\frac {{\delta}^{3}{w}^{3}\beta}{{\varepsilon}^{2}}}-{ \frac
{55460160}{30371}}\,{\frac {{\delta}^{5}{w}^{3}}{{\varepsilon}^{3}} }+\\
\\+\frac12\,{C_0}^{2}+{\frac {660240}{2761}}\,{\frac {{\delta}^{3}{w}^
{2}\beta}{{\varepsilon}^{2}}}-{\frac {25200}{30371}}\,{\frac {{\delta}^{5
}{w}^{2}}{{\varepsilon}^{3}}}-{\frac {1260}{251}}\,{\frac {{\beta}^{2}
\delta\,w}{\varepsilon}}+{\frac {12600}{30371}}\,{\frac {\beta\,{\delta}^
{3}w}{{\varepsilon}^{2}}}
\end{gathered}\end{equation}

Substituting solution \eqref{3.15} into \eqref{3.14} and taking into account that
$\alpha= \alpha_i=w_i \delta/\varepsilon$ $(i=1, ..., 6)$ we have different solutions of
Eq.\eqref{1.1} in the form of  solitary waves.

\section{Exact periodic solutions of Eq.(1.1).}

We can see that solutions of Eq\eqref{1.1} have fivth degree singularity and one can also
look for exact solution of Eq.\eqref{1.1} in the form

\begin{equation}
\label{4.1}y \left( z \right) =B_1+ B_2 R ( z ) + B_3R_z + B_4 R^{2}+B_5RR_z
\end{equation}
where $B_k \,\,(k=1, ..., 5)$ are constants and $R=R(z)$ is a second degree singularity
solution of the elliptic function equation

\begin{equation}
\label{4.2}R^2_z =-2R^3 +aR^2 +2bR +d
\end{equation}

From Eq.\eqref{4.2} we get that $R(z)$ satisfies also to equations

\begin{equation}
\begin{gathered}
\label{4.3} R_{zz} =-3R^2 +aR+b\\
\\
R_{zzz} =-6\,R  R_z +a R_z \\
\\
R_{zzzz}=30\,  R  ^{3}-15\,a R    ^{2}-
18\,bR  -6\,d+{a}^{2}R  +ab \\
\\
R_{zzzzz}=90\,  R  ^{2}R_z -30\,aR
 R_z -18\,b R_z +{a}^{2}R_z \\
\\
R_{zzzzzz}=-630\,  R   ^{4}+420\,a  R   ^{3}+ 504\,b  R  ^{2}+180\,R  d-\\
 \\-63\,{a}^{2}  R  ^{2}-108\,ab
R  -30\,ab-18\,{b}^{2}+{a}^{3}R  +{a}^ {2}b
\end{gathered}
\end{equation}

Substituting \eqref{4.1} into Eq.\eqref{1.1} and taking into account formulas \eqref{4.2}
and \eqref{4.3} we find

\begin{equation}
\label{4.4}B_4=0,\quad B_2=0,\quad  B_1= C_0,\quad  B_5=-3780\,\varepsilon,\quad
B_3=630\varepsilon\,a+{\frac {630}{11}}\,\delta
\end{equation}

\begin{equation}
\label{4.5}\beta={\frac {10}{121}}\,{\frac {{\delta}^{2}}{\varepsilon}}
\end{equation}

As this takes place parameters $b$ and $d$ in Eq.\eqref{4.1} take two values

\begin{equation}
\begin{gathered}
\label{4.6} b_{1,2}=-\frac1{12} {a}^{2}+{\frac {1}{1452}}\,{\frac {{\delta}^{2}}{{
\varepsilon}^{2}}}\pm {\frac {1}{5082}}\,{\frac {\sqrt {21}{\delta}^{2}}{{
\varepsilon}^{2}}}
\end{gathered}
\end{equation}

\begin{equation}
\begin{gathered}
\label{4.7}d_{1,2}={\frac {1}{108}}\,{a}^{3}+{\frac {13}{359370}}\,{\frac {{
\delta}^{3}}{{\varepsilon}^{3}}}\pm{\frac {1}{119790}}\,{\frac {\sqrt {21}{
\delta}^{3}}{{\varepsilon}^{3}}}-{\frac {1}{4356}}\,{\frac {a{\delta}^{2}
}{{\varepsilon}^{2}}}\mp{\frac {1}{15246}}\,{\frac {a\sqrt {21}{\delta}^{2}
}{{\varepsilon}^{2}}}
\end{gathered}
\end{equation}

Constant $C_1$ in Eq.\eqref{2.1} in this case has two values too

\begin{equation}
\begin{gathered}
\label{4.8}C^{(1,2)}_1=-{\frac {10854}{161051}}\,{\frac {{\delta}^{5}}{{\varepsilon}^{
3}}}+\frac12{C_0}^{2}\mp{\frac {2484}{161051}}\,{\frac {\sqrt {21}{
\delta}^{5}}{{\varepsilon}^{3}}}
\end{gathered}
\end{equation}

Using \eqref{4.4} and \eqref{4.5} we obtain as resultant expression for the solution
$y(z)$ in the form of periodic waves.

\begin{equation}
\begin{gathered}
\label{4.9} y(z)=C_0 + 630 \left(\varepsilon a +\frac\delta {11} -6\varepsilon R\right)
R_z
\end{gathered}
\end{equation}
where $R=R(z)$ is a solution of the following equations

\begin{equation}\begin{gathered}
\label{4.10} R^2_z=-2\,  R
   ^{3}+a  R ^
{2}-\frac16{a}^{2}R  +{\frac {1}{726}}{\frac {R
  {\delta}^{2}}{{\varepsilon}^{2}}}\pm
 {\frac {1}{2541}}{
\frac {R  \sqrt {21}{\delta}^{2}}{{\varepsilon}^{2}}}+\\ \\{ \frac
{1}{108}}\,{a}^{3}+{\frac {13}{359370}}\,{\frac {{\delta}^{3}}{{
\varepsilon}^{3}}}\pm{\frac {1}{119790}}\,{\frac {\sqrt {21}{\delta}^{3}}{{
\varepsilon}^{3}}}-{\frac {1}{4356}}\,{\frac {a{\delta}^{2}}{{\varepsilon}^{
2}}}\mp{\frac {1}{15246}}\,{\frac {a\sqrt {21}{\delta}^{2}}{{\varepsilon}^{ 2}}}
\end{gathered}\end{equation}

Assuming that $R_1,R_2$ and $R_3$ with $R_1\geq R_2 \geq R_3$ real roots of equations

\begin{equation}
\begin{gathered}
\label{4.11} 2R^3 -aR^2 +\left(\frac16 a^2 - \frac{\delta^2}{726\varepsilon^2} \mp
\frac{\delta^2 \sqrt{21}}{2541\varepsilon^2}\right)R-\frac1{108} a^3-\\ \\
-\frac{13}{359379} \frac{\delta^3}{\varepsilon^3} \mp
\frac{\delta^3\sqrt{21}}{119790\varepsilon^3} +\frac{a\delta^2}{4356\varepsilon^2}
\pm\frac {a\delta^2\sqrt{21}}{15246\varepsilon^2}=0
\end{gathered}
\end{equation}

We have solutions of Eq.\eqref{4.10} in the form

\begin{equation}
\begin{gathered}
\label{4.12} R(z) =R_2 + (R_1-R_2) \textup{cn}^2(z\sqrt{R_1-R_2},S),\quad
S^2=\frac{R_1-R_2}{R_1-R_3}
\end{gathered}
\end{equation}

 Thus Eq.\eqref{1.1} have a few exact solutions at different values of
equation parameters. These solution are  solitary and periodic waves and they are
determined by the formulas \eqref{3.14} and \eqref{4.9}. We hope these solutions can be
useful for test of the numerical simulations of soft-mode turbulence.

This work was supported by the International Science and Technology Center under the
project 1379-2.


\newpage
\begin{thebibliography}{99}

\bibitem{1} M.I. Tribelsky and K. Tsuboi, Phys. Rev. Lett. 76, 1631 (1996)

\bibitem{2} Hao-wen Xi, Raul Toral, J.D. Gunton and Michael I. Tribelsky, Phys. Rev E., 62,1 (2000) 17-20

\bibitem{3} Raul Toral, Guoming Xiong, J. D. Gunton and Hao-wen Xi, J.Phys.A. Math. Gen.
v. 36 (2003) 1323-1335

\bibitem{4} N.A. Kudryashov, Mathematical simulation, vol 1, No 6 (1989), 55 (in Russian)

\bibitem{5} L.A. Beresnev and V.N. Nikolaevskiy Physyca D, 66 (1993) 1-6

\bibitem{6} R. Conte and M. Musette, J. Phys. A., 22 (1989), 169-177

\bibitem{7} N.A. Kudryashov, Mathenatical simulation, vol 1, No 9 (1989) 151-158 (in
Russian)

\bibitem{8} N.A. Kudryashov, Journal of Applied Mathematics and Mekhanics, 52, 3 (1988)
361-365

\bibitem{9} N.A. Kudryashov, Phys Lett. A., 147 (1990) 287-291

\bibitem{10} N.A. Kudryashov, Phys Lett. A., 155 (1991) 269-275

\bibitem{11} N.A. Kudryashov, and E.D. Zargaryan, J. Phys. A. Math. and Gen 29 (1996) 8067-8077

\bibitem{12} M. Musette and R. Conte, Physica D, 181 (2003) 70-79

\bibitem{13} Z. Fu, S. Liu, S. Liu, Q. Zhao Phys. Lett. A. 290 (2001) 72-76

\bibitem{14} Z. Fu, S. Liu, S. Liu Phys. Lett. A., 299 (2002) 507-512

\bibitem{15} N.G. Berloff and L.M. Howard, Studies in Applied Mathematics, 100 (1998)
195-213

\bibitem{16} Z.Y. Yan, Chaos Solitons Fractuls, 15, 3 (2003) 575-583

\bibitem{17} E.G. Fan, Comput. math. appl., 43, 6-7 (2002) 671-680

\bibitem{18} E.Fan, Nuovo Cimento B, 116, 12 (2001) 1385-1393

\bibitem{19} S.A. Elwakil, S.K. Ellabany, M.A. Zahran, et al., Phys. Lett. A., 299, 2-3
(2002) 179-188

\bibitem{20} R.X. Yao, Z.B. Li, Chinese Phys, 11, 9 (2002) 864-868

\bibitem{21} J. Weiss, M. Tabor and G. Carnevalle, J. Math. Phys. 24, 3 (1983) 522-526

\bibitem{22} N.A. Kudryashov, Analytical theary of nonlinear diffeential equations,
Moscow-Igevsk, IKI (2003) 360 p


\end{thebibliography}
\end{document}